\documentclass[%
 prl,
rsi,%
amsmath,amssymb,
reprint,%
]{revtex4-1}

\usepackage{graphicx}
\usepackage{dcolumn}

\begin{document}

\author{Victor Despr{\'e}}
\email{victor.despre@pci.uni-heidelberg.de}
\author{Alexander I. Kuleff}
\email{alexander.kuleff@pci.uni-heidelberg.de}
\affiliation{Theoretische Chemie, PCI, Universit{\"a}t Heidelberg, Im Neuenheimer Feld 229, D-69120 Heidelberg, Germany}

\title{Correlation-Driven Charge Migration as Initial Step in the Dynamics\\in Correlation Bands}

\date{\today}

\begin{abstract}
We present dynamics calculations showing how electron-correlation-driven charge migration occurring in the correlation band of ionized molecules can lead to a redistribution of the charge increasing the stability of the system. These calculations offer an interpretation of recent experimental results obtained for adenine. We discuss the implications of the mechanism for the development of attochemistry and how it can be understood in the context of the ultrafast, non-adiabatic relaxation taking place in highly-excited molecular cations.
\end{abstract}

\maketitle
The development of ultrafast technology allows nowadays to study electron dynamics in their intrinsic time scale \cite{nisoli2017attosecond}. The observation of pure electron dynamics occurring after photo-ionization of molecular systems \cite{calegari2014ultrafast,kraus2015measurement,lara2018attosecond} is a first step towards the realization of attochemistry, namely the possibility to control the reactivity of a molecular system by acting only on its electronic degrees of freedom. These dynamics, referred to as charge migration, have been intensively studied theoretically in the last two decades (see, e.g., Ref.~\cite{remacle1998charge,cederbaum1999ultrafast,hennig2005electron,yudin2005attosecond,kuleff2005multielectron,barth2006periodic,kuleff2010ultrafast,kus2013pump,lara-astiaso2016decoherence,sun2017nuclear,mauger2018signature,despre2019size,perfetto2020ultrafast,schwickert2020electronic,folorunso2021molecular}), showing that charge migration is a rich phenomenon, with many facets that are rather characteristic for the investigated molecule. A mechanism of particular interest in this context is the so-called correlation-driven charge migration \cite{cederbaum1999ultrafast,kuleff2014ultrafast}, in which the hole-charge created by ionization can migrate throughout the molecule on a few-femtosecond time scale. These charge dynamics appear as a result of the population of specific coherent superpositions of electronic states upon ionization, determined by the electron-correlation effects in the molecule. Despite of the great interest, the experimental study of the correlation-driven charge migration is still very challenging and the phenomenon has not been directly observed yet. One of the reasons is that even at fixed nuclei, the zero-point spread of the nuclear wave packet causes an electronic dephasing that puts fast to an end the pure electronic charge migration \cite{despre2015attosecond,vacher2015electron}. Moreover, the coupling to the nuclear motion may destroy the initially created electronic coherence extremely fast. Decoherence times of less than 3~fs have been predicted for certain molecules and electronic states \cite{vacher2017electron,arnold2017electronic}. It has to be noted, however, that longer lived electronic coherences have also been predicted for polyatomics \cite{despre2015attosecond,despre2018charge}, offering suitable systems to study the correlation-driven charge migration phenomenon.

Very recently, in a pioneering experiment, Calegari and co-workers where able to identify a sub 3~fs pure electron dynamics in adenine \cite{maansson2021real}. This dynamics triggered by an XUV ionization is observed through a positive delay in the appearance of the parent-dication signal after a second ionization of the adenine with an NIR pulse in a pump-probe scheme. As the analysis showed that only two NIR photons are absorbed, the probed dynamics have to take place a few eV below the double-ionization threshold, i.e., in the correlation band of adenine. The electron dynamics thus causes a delay in the onset of the non-adiabatic relaxation in the correlation band of the system. The characteristic correlation-band relaxation dynamics were recently identified in a series of polycyclic aromatic hydrocarbon molecules \cite{Herve2020Ultrafast} and allowed also to interpret previously published results \cite{belshaw2012observation,marciniak2015xuv,marciniak2018ultrafast}. Correlation bands \cite{deleuze1996formation} exist in all molecular systems of sufficient size and appear in their ionization spectra as a large increase of the density of states just below the double-ionization threshold, due to the electron-correlation-induced breakdown of the molecular-orbital picture \cite{cederbaum1986correlation}. It was shown \cite{Herve2020Ultrafast} that the relaxation in correlation bands follows a characteristic scaling low for a given family of molecular systems of increasing size, competing with the simultaneously occurring redistribution of the energy among the different vibrational degrees of freedom \cite{boyer2021ultrafast}. In the context of attochemistry, the identification of pure electron dynamics preceding the universal non-adiabatic relaxation in the correlation band is very important, as such an event sequence may offer a possibility to control the overall relaxation mechanism of highly-excited molecular cations.

In Ref.~\cite{maansson2021real}, the authors interpreted the observed electron dynamics as a frustrated Auger decay \cite{bagus2004anew} (referred to as shake-up process in Ref.~\cite{maansson2021real}). In their interpretation, the XUV pump ionizes the neutral adenine, creating a cationic species with a vacancy in an inner-valence orbital. The system then undergoes a frustrated decay process in which a valence electron fills the initial vacancy and another one is promoted to a virtual orbital that the authors identify as LUMO$+6$. It is supposed that the NIR probe pulse can then ionize the electron from the LUMO$+6$. The delay in the appearance of the dicationic signal was therefore interpreted as the time scale of this frustrated decay process, which opens the possibility for ionization with two IR photons. While formally possible, such a mechanism seems unlikely to be the reason for the observed dynamics. First, as no energy dissipation takes place, both the state before and after the frustrated Auger decay have the same energy and thus can be ionized by the IR probe. The delay, therefore, has to come from an increase in the ionization cross section during the process. The probability to ionize the highly excited and thus very diffused LUMO$+6$ electron, however, is expected to be smaller than that to eject a more localized, valence electron \cite{chang1982photoionization}, suggesting that the ionization cross section will rather decrease during the frustrated decay. Second, in the energy range considered, i.e., a few eV below the double-ionization threshold, the correlation band forms and the notion of individual states and electron configurations ceases to be an adequate description of the highly excited system. Due to the strong many-body effects there, the removal of an electron populates a quasi-continuum of states that involve a large number of particle-hole excitations. 

In this letter we present an alternative interpretation of the experimental results of Ref.~\cite{maansson2021real}, namely that the observed delay in the appearance of the dicationic signal is due to the ultrafast correlation-driven charge migration, resulting from the population of the correlation band of adenine. We show that the population of the part of the correlation band that is probed in the experiment is formed by ionization out of three $\sigma$-orbitals, and the subsequent sub 3~fs charge migration represents a redistribution of the charge from a hole with $\sigma$ to a hole with $\pi$ character. It is the transfer of positive charge from the bonds to an out-of-molecular-plane $\pi$ structure that explains the higher stability of the system after the charge migration. We will also discuss how this mechanism can be interpreted in the context of electron-phonon scattering model used to describe the non-adiabaic dynamics in correlation bands \cite{Herve2020Ultrafast}. 

\begin{figure}
    \centering
    \includegraphics[scale = 0.4]{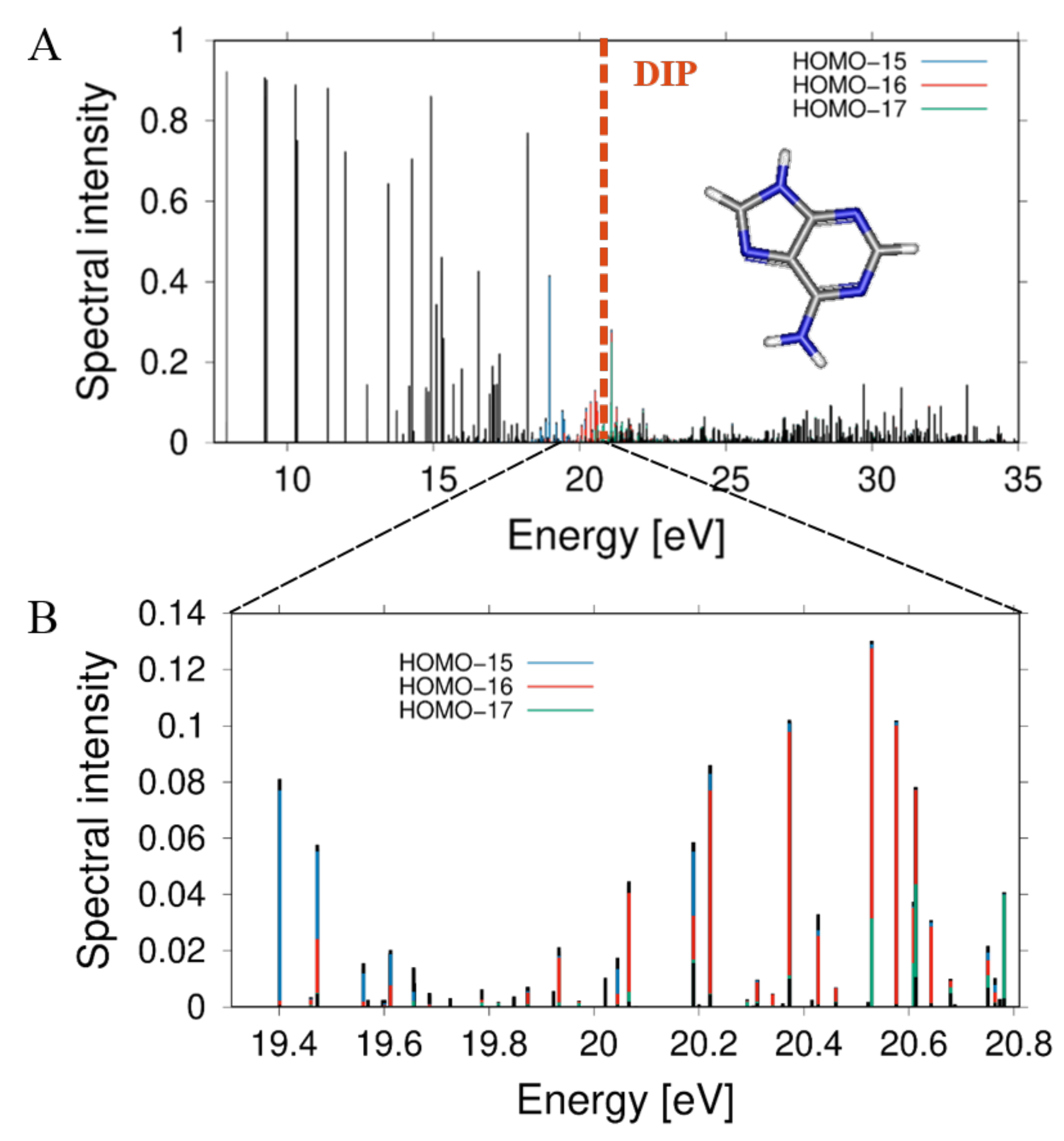}
    \caption{(A) Ionization spectrum of adenine computed with ADC(3). The dashed orange line, located at at 20.8~eV, shows the position of the double-ionization potential (DIP) as computed with ADC(2)/cc-pVDZ level of theory. (B) Zoom of the correlation band region. The contributions of the one-hole configurations corresponding to the removal of an electron from HOMO$-15$, HOMO$-16$, and HOMO$-17$ are depicted with cyan, red, and green, respectively.}
    \label{fig:spec}
\end{figure}

The starting point for understanding the ultrafast dynamics induced by the pump pulse is the analysis of the ionization spectrum of adenine. To compute the latter we used the non-Dyson ADC(3) method \cite{schirmer1998non} employing cc-pVDZ basis set. ADC, standing for Algebraic Diagrammatic Construction, is a Green's function based perturbative method that provides a high-level \textit{ab initio} description of the electronic structure of an ion and has the advantage to provide all cationic states of the system in a single calculation. We used the third-order ADC scheme [ADC(3)], in which all possible one-hole (1h) and two-holes-one-particle (2h1p) configurations that can be formed within the molecular-orbital basis used are taken into account. This method has been shown to reproduce well the experimental photoelectron spectrum of adenine \cite{trofimov2005photoelectron}. The ADC(3)/cc-pVDZ ionization spectrum of adenine is presented in Fig.~\ref{fig:spec}. Each line in the spectrum represents a cationic eigenstate and is located at the corresponding ionization potential. Its height, or spectral intensity, is directly related to the probability to populate the corresponding state, and is given by the 1h part of the state (the total weight of all contributing 1h configurations). As the 2h1p configurations describe excitations on top of the the removal of a particular electron, their weight is a measure of the correlation effects contributing to the state.

The spectrum of adenine exhibits the typical behavior, with states dominated by 1h configurations at low energy, followed by a region where the multi-electronic effects become stronger and many satellite structures appear (between approximately 14 and 18~eV). Above $\sim 18$~eV, we observe a complete breakdown of the molecular orbital picture, and the spectrum becomes a quasi-continuum of states dominated by multi-excitations -- the correlation band of the system. Although the XUV pulse used in Ref.~\cite{maansson2021real} can populate all the sates below $\sim 32$~eV, the second ionization performed by the IR probe restricts the observed dynamics to the correlation band. The double-ionization threshold, denoted in Fig.~\ref{fig:spec}~A by a dashed orange line, was computed with the particle-particle ADC(2) method \cite{schirmer1984higher} using again cc-pVDZ basis set and lies at 20.8~eV (we note, however, that the value used in the analysis of Ref.~\cite{maansson2021real} is 22.2~eV).

Taking a closer look at the correlation-band region (see Fig.~\ref{fig:spec}~B), directly below the double-ionization potential, we see that it is dominated by states populated by ionization out of HOMO$-16$ with a smaller contribution from states stemming from ionization of HOMO$-15$ and HOMO$-17$. All these three orbitals are $\sigma$ orbitals. In what follows, we will concentrate on the pure electron dynamics taking place after the ionization of HOMO$-16$, noting that the charge migration triggered by the ionization out of HOMO$-15$ and HOMO$-17$ is very similar and can be found in the Supplemental Material (SM).

The evolution of the electronic cloud following the removal of a HOMO$-16$ electron can be traced with the help of the multielectron wave-packet propagation method \cite{kuleff2005multielectron}, based on a direct propagation of the initially created electronic wave packet with the full ADC(3) cationic Hamiltonian via the short iterative Lanczos technique \cite{park1986unitary}. A convenient way to visualize and analyze the dynamics is to compute the hole density, defined as the deference between the electron density of the neutral system, $\rho_0(\vec{r})$, and the time-dependent electron density of the cation, $\rho_i(\vec{r},t)$, \cite{cederbaum1999ultrafast}
\begin{equation}
Q(\vec{r},t) = \underbrace{\langle\Psi_{0}|\hat{\rho}|\Psi_{0}\rangle}_{\rho_0 (\vec{r})} - \underbrace{\langle\Phi_{i}(t)|\hat{\rho}|\Phi_{i}(t)\rangle}_{\rho_{i}(\vec{r}, t)},
\end{equation}
where $|\Psi_{0}\rangle$ is the neutral ground state, $|\Phi_{i}(t)\rangle$ is the propagated wave packet created by the ionization at $t=0$, and $\hat{\rho}$ is the electronic density operator. Using the molecular orbitals of the neutral system as a basis, one can obtain the following expression for the hole density \cite{breidbach2003migration}
\begin{equation}
    Q( \vec{r},t) = \sum_{p} |\tilde{\varphi}_{p}( \vec{r},t)|^2 \tilde{n}_{p}(t),
\end{equation}
where $\tilde{\varphi}_{p}( \vec{r},t)$ are the natural charge orbitals (given at each time point by different linear combinations of the neutral MOs) and $\tilde{n}_{p}(t)$ their hole-occupation numbers. Further details and the derivation of the above expression can be found in Ref.~\cite{breidbach2003migration,breidbach2007migration,kuleff2005multielectron}.

\begin{figure}
    \centering
    \includegraphics[scale = 0.33]{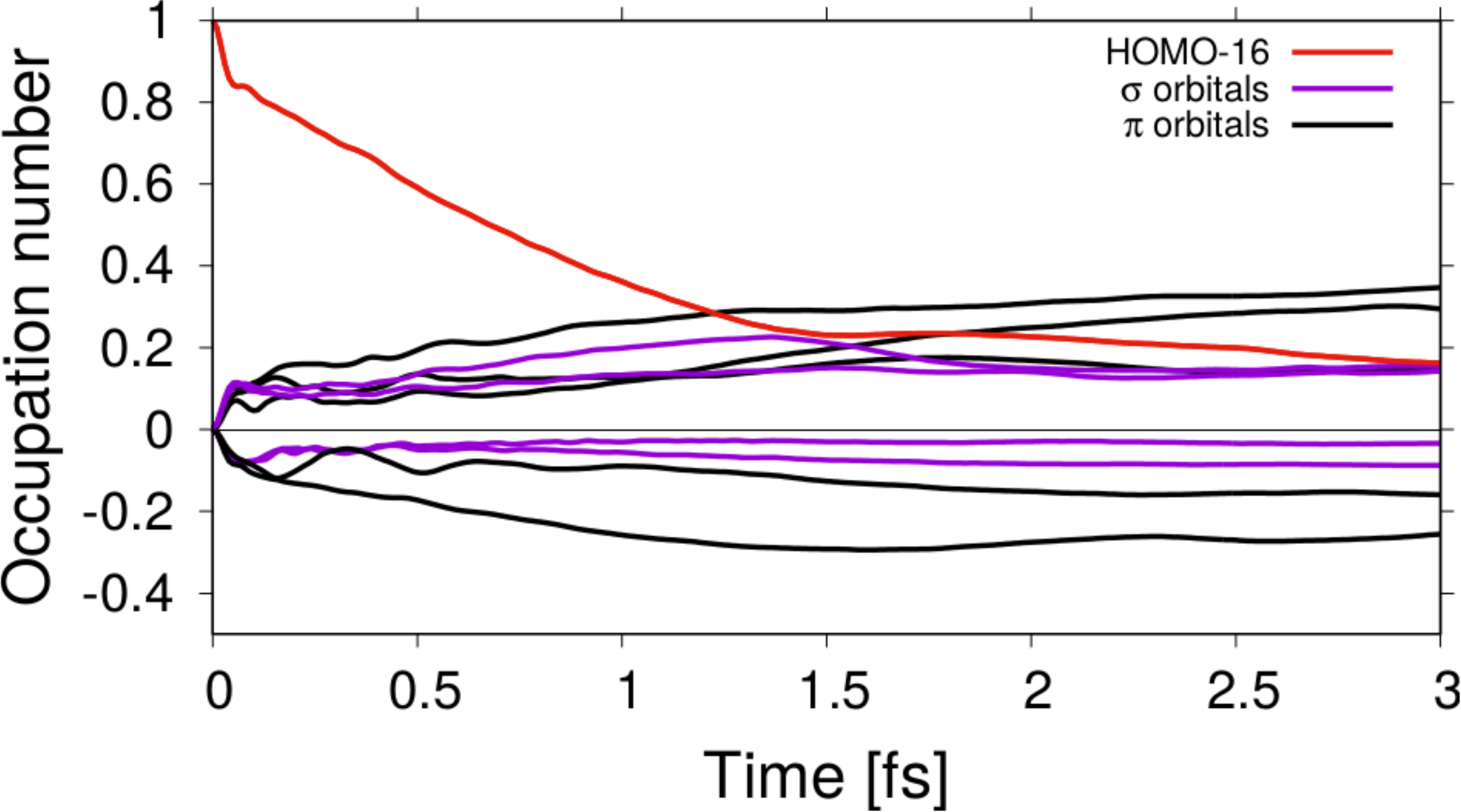}
    \caption{Evolution of the hole occupations following the removal of an electron from HOMO$-16$. A negative hole occupation means that an electron is promoted to a virtual orbital. We see that for less than 3~fs, the initial hole (red curve) migrates preferentially to $\pi$ orbitals (depicted in black).}
    \label{fig:hole_occupation}
\end{figure}

The evolution of the most important hole-occupation numbers following the ionization of HOMO$-16$ is shown in Fig.~\ref{fig:hole_occupation} (the hole-occupation numbers following the ionization of HOMO$-17$ and HOMO$-15$ can be found in the SM, Fig.~S1 and Fig.~S3, respectively). It is important to note that removing an electron from HOMO$-16$ does not mean that a single cationic state is populated. All states, energetically accessible by the pump pulse, with 1h configuration describing an electron missing in HOMO$-16$ get populated with the corresponding weights (depicted in red in Fig.~\ref{fig:spec}). An electronic wave packet is thus created by the ionization. We see from Fig.~\ref{fig:hole_occupation} that at time $t = 0$ the hole is in the initially ionized HOMO$-16$ orbital. As time progresses, the initial hole is filled (the HOMO$-16$ hole occupation decreases) and the hole population is transferred preferentially to orbitals with $\pi$ character (depicted in black in Fig.~\ref{fig:hole_occupation}). We note that a negative hole-occupation number means an electron in a virtual natural charge orbital and thus the hole occupations evolving symmetrically around zero in Fig.~\ref{fig:hole_occupation} describe electronic excitations. The overall dynamics has a time scale of less that 3~fs, in good accord with the observations reported in Ref.~\cite{maansson2021real}.

\begin{figure*}
    \centering
    \includegraphics[scale = 0.55]{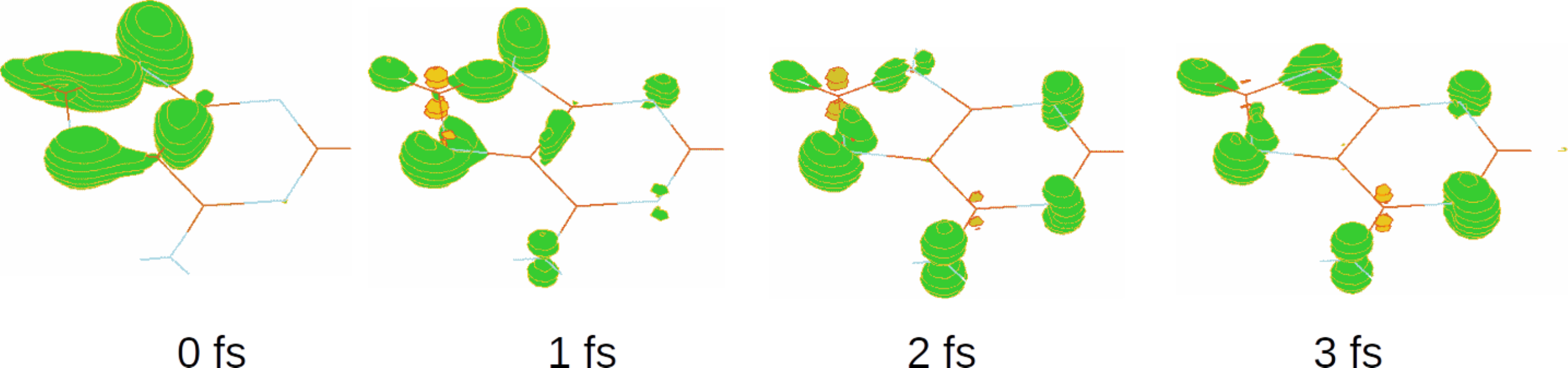}
    \caption{Snapshot of the hole density at 0, 1, 2, and 3~fs following the removal of an electron from HOMO$-16$. Lack of electron density (hole) is depicted in green, while areas with excess of electron density are depicted in orange.}
    \label{fig:density}
\end{figure*}

To better understand the spatial evolution of the charge during the process, we show in  Fig.~\ref{fig:density} snapshots of the hole density following ionization out of HOMO$-16$ (similar hole density dynamics is observed after the ionization of HOMO$-17$ and HOMO$-15$, see Fig.~S2 and Fig.~S4, respectively, in the SM). At $t = 0$ the hole is located on the pentagonal ring and, as a consequence of the $\sigma$ character of HOMO$-16$, is localized on several bonds. As time progresses, the hole gets delocalized and spreads over the molecule, changing its character from $\sigma$ to $\pi$. As a result of the latter, the positive charge is no longer concentrated on the bonds, but gets localized on the atoms, preferentially outside the molecular plane.

Let us now discuss how the experimental results of Ref.~\cite{maansson2021real} can be interpreted with the help of these correlation-driven charge migration dynamics. The XUV-pump pulse will ionize the system populating all the states below $\sim 32$~eV, but the second ionization by IR-probe restricts the observed dynamics to the spectral region spreading just a few eV below the double-ionization threshold. This part of the correlation band of adenine is formed by the ionization out of three orbitals, namely HOMO$-17$ to HOMO$-15$. The population of the resulting quasicontinuum of cationic states triggers a sub 3~fs correlation-driven charge-migration dynamics, in which the hole, initially located mostly on the bonds and thus weakening them, redistributes throughout the system such that at the end of the process it is mostly localized outside the molecular skeleton. This ultrafast charge redistribution, therefore, stabilizes the cation. If the molecule is further ionized before this stabilization process is completed, the dication is more likely to break apart and thus to contribute to the cationic-fragments signals observed. On the contrary, a second ionization performed after the cationic stabilization will increase the probability to observe a stable dicationic system and thus the parent-dication signal will start to increase. The measured sub 3~fs delay in the appearance of the parent-dication yield is, therefore, a signature of the correlation-driven charge migration taking place in the correlation band of adenine. 

Let us now discuss our results in a broader context. While the delay observed in the dicationic signal of adenine could be specific for this molecule, the correlation-driven charge migration triggered by the population of the correlation band is a general effect. Indeed, correlation bands are formed in the cationc spectra of every polyatomic molecule and thus a removal of an electron from deep inner-valence orbitals will inevitably lead to the population of a quasi-continuum of states spread over, sometimes, several eV. Such an electronic wave packet will evolve on an ultrashort time scale (a few fs) redistributing the charge throughout the system and preparing it for the follow-up non-adiabatic dynamics. The correlation-driven charge-migration process will, therefore, serve as an initial step in the relaxation in the correlation band. Extending the analogy between correlation bands in molecules and energy bands in solids introduced in Ref.~\cite{Herve2020Ultrafast} and allowing to describe the non-adibatic relaxation in correlation bands with an electron-phonon scattering model, the correlation-driven charge migration can be seen as an equivalent of the electron cooling occurring through electron-electron collisions in solids. In both cases, the energy is first localized on a specific part of the system and then, due only to the electron correlation, efficiently spreads over the system. This offers a new perspective in the understanding of the physics of correlation bands and the relaxation of highly excited molecular cations. 

In the end, we would like to touch upon the possible consequences of such a sequence of ultrafast processes initiated by the ionization of a molecule with an XUV light. Following the concept of attochemistry \cite{palacios2020quantum}, the possibility to exert control over the initial, pure electronic charge-redistribution step can be used to modulate the follow-up relaxation dynamics and thus the chemical reactivity of the molecule, eventually influencing the fragmentation pattern. Efficient control over charge migration dynamics has already been demonstrated theoretically in the outer-valence \cite{golubev2015control,golubev2017quantum}, where typically only several states participate. The extension of these techniques to dense spectral regions and the development of new ones might offer interesting perspectives in quantum control of photoinduced chemistry and a deeper understanding of the radiation damage in (bio)matter. 

\begin{acknowledgments}

The authors acknowledge financial support from the DFG through the QUTIF priority programme.

\end{acknowledgments}


\bibliography{adenine_CB}

\end{document}